\documentclass[sigconf,9pt,nonacm]{acmart}

\usepackage{booktabs}
\usepackage{algorithm}
\usepackage{algpseudocode}
\usepackage{graphicx}
\usepackage{textcomp}
\usepackage{xcolor}
\usepackage{url}
\usepackage{cuted}
\usepackage{capt-of}
\usepackage{seqsplit}
\usepackage{enumitem}

\usepackage[caption=false,font=footnotesize]{subfig}

\usepackage{tikz}
\newcommand{\popnum}[1]{%
\tikz[baseline=(n.base)]\node
[circle, fill=black, text=white, inner sep=1.6pt, font=\footnotesize\bfseries] (n) {#1};%
}

\usepackage[framemethod=tikz]{mdframed}
\newmdenv[
  linewidth=0.8pt,
  linecolor=black!50,
  backgroundcolor=gray!20,
  roundcorner=12pt,
  skipabove=6pt,
  skipbelow=6pt,
  innerleftmargin=10pt,
  innerrightmargin=10pt,
  innertopmargin=10pt,
  innerbottommargin=10pt
]{takeawaybox}

\setcopyright{none}
\setkeys{acmart.cls}{balance=false}

\begin{document}

\title{Where Does My Call Go? Measuring Google Meet, Zoom, and Microsoft Teams on Starlink}

\author{Kousar Malekinasab}
\affiliation{%
  \institution{University of Victoria}
  \city{Victoria}
  \state{BC}
  \country{Canada}}
\email{kovvy@uvic.ca}

\author{Mostafa Abdollahi}
\affiliation{%
  \institution{University of Victoria}
  \city{Victoria}
  \state{BC}
  \country{Canada}}
\email{abdolahi68@uvic.ca}

\author{Jinwei Zhao}
\affiliation{%
  \institution{University of Victoria}
  \city{Victoria}
  \state{BC}
  \country{Canada}}
\email{clarkzjw@uvic.ca}

\author{Jianping Pan}
\affiliation{%
  \institution{University of Victoria}
  \city{Victoria}
  \state{BC}
  \country{Canada}}
\email{pan@uvic.ca}

\begin{abstract}
Low-Earth-Orbit (LEO) satellite networks are expanding broadband access, but major Real-Time Communication (RTC) platforms, including Google Meet, Zoom, and Microsoft Teams, still rely on service-point selection mechanisms that appear to be largely shaped by terrestrial-network assumptions. As a result, they may not consistently account for LEO-specific dynamics such as satellite topology and Point-of-Presence (PoP) changes, especially during in-flight connectivity. To characterize RTC service-point selection over Starlink, we measure RTC performance using a Canada-wide testbed and more than 24 hours of in-flight experiments. Our results show clear platform differences. Google Meet tends to align more closely with the serving PoP, consistent with broader observed edge coverage. Zoom follows a regional service-point selection pattern, while Microsoft Teams often remains tied to fixed service points. Moreover, we show that in-flight PoP handovers can shift traffic egress points mid-session, and that RTC platforms respond differently to these changes. In some cases, this interaction leads to suboptimal service-point selections and RTT inflation. Overall, our findings suggest that LEO connectivity can expose mismatches between Starlink egress dynamics and RTC service-point selection, motivating further measurement and LEO-optimized designs for service placement and selection.
\end{abstract}

\ccsdesc[500]{Networks~Network measurement}
\ccsdesc[300]{Networks~Network performance analysis}

\keywords{Real-Time Communication, LEO satellite networks, Starlink, WebRTC, network measurement}

\maketitle

\section{Introduction}

\par Real-Time Communication (RTC) has become a fundamental Internet service and is now widely used in online meetings, virtual classrooms, telehealth, and live collaborative applications. At Internet scale, major RTC platforms such as Google Meet, Zoom, and Microsoft Teams serve hundreds of millions of users~\cite{zoom300m2021, Soltero2020Meet, microsoftteams2023ecosystem}. At this scale, WebRTC~\cite{alvestrand2021rfc} has become a key framework for real-time audio, video, and data communication and is widely adopted by modern communication platforms. To deliver low-latency communication at this scale, providers rely on geographically distributed service infrastructures, including media edges, relay servers, and private backbones. In terrestrial networks, such designs generally improve performance by bringing traffic into the provider network close to the user and carrying media over well-provisioned backbone paths.

\par However, it remains unclear how these designs perform over LEO satellite networks such as Starlink. In this setting, end-to-end performance depends not only on user and provider locations, but also on the Starlink Point-of-Presence (PoP) through which traffic exits the satellite network~\cite{pan2023measuring}. This dependency is especially important in high-mobility scenarios such as
commercial flights, where emerging LEO-based in-flight connectivity is making
interactive and media-rich applications increasingly feasible, such as video conferencing and real-time streaming
\cite{BA_RTC,Qatar_Starlink}. In such settings, PoP handovers and the changing
user--PoP distance can alter routing paths and latency even if the application
keeps the same service point. Thus, server-selection strategies commonly used in terrestrial settings may not fully account for LEO access dynamics. Although recent studies examine the performance of real-time applications such as Zoom over Starlink~\cite{mohan2024multifaceted,fang2025streaming,zhao2023realtime}, they mainly focus on client-side metrics and do not analyze how RTC providers select media servers or service edges based on PoP location, user distribution, or mobility. A systematic comparison of Google Meet, Zoom, and Microsoft Teams for stationary and in-flight Starlink users, and the resulting impact on routing and latency, is still missing. 

\par In this paper, we present a measurement study of browser-based RTC (i.e., WebRTC) over the Starlink network using Google Meet, Zoom, and Microsoft Teams. Our study covers both stationary and in-flight users. For stationary users, we examine multiple RTC session scenarios by varying the Starlink PoP associations and the number of participants. These scenarios include two-party sessions in which participants are connected through either the same Starlink PoP or different PoPs, as well as multi-party sessions with more than two participants. Our measurements use the Coast-to-Coast-to-Coast Low-Earth-Orbit Satellite Networks (CCCLEO) testbed in Canada\footnote{\url{https://oac.uvic.ca/cccleo}}. As shown in Figure~\ref{fig:figLeo}, we select seven Starlink nodes located in Victoria, Whitehorse, Saskatoon, Thompson, Iqaluit, Fredericton, and St.~John's. For in-flight users, we analyze six Starlink-enabled commercial flights, totaling more than 24 hours of traces, to capture user in-flight mobility, PoP handovers, and changes in user--PoP distance during live sessions. Using a containerized Chromium-based framework with automated session control and WebRTC telemetry extraction, we analyze service-point selection, routing paths, and RTT behavior across these scenarios. Our measurements reveal clear platform-specific strategies:

\begin{figure}[!t]
    \centering
    \includegraphics[width=\linewidth]{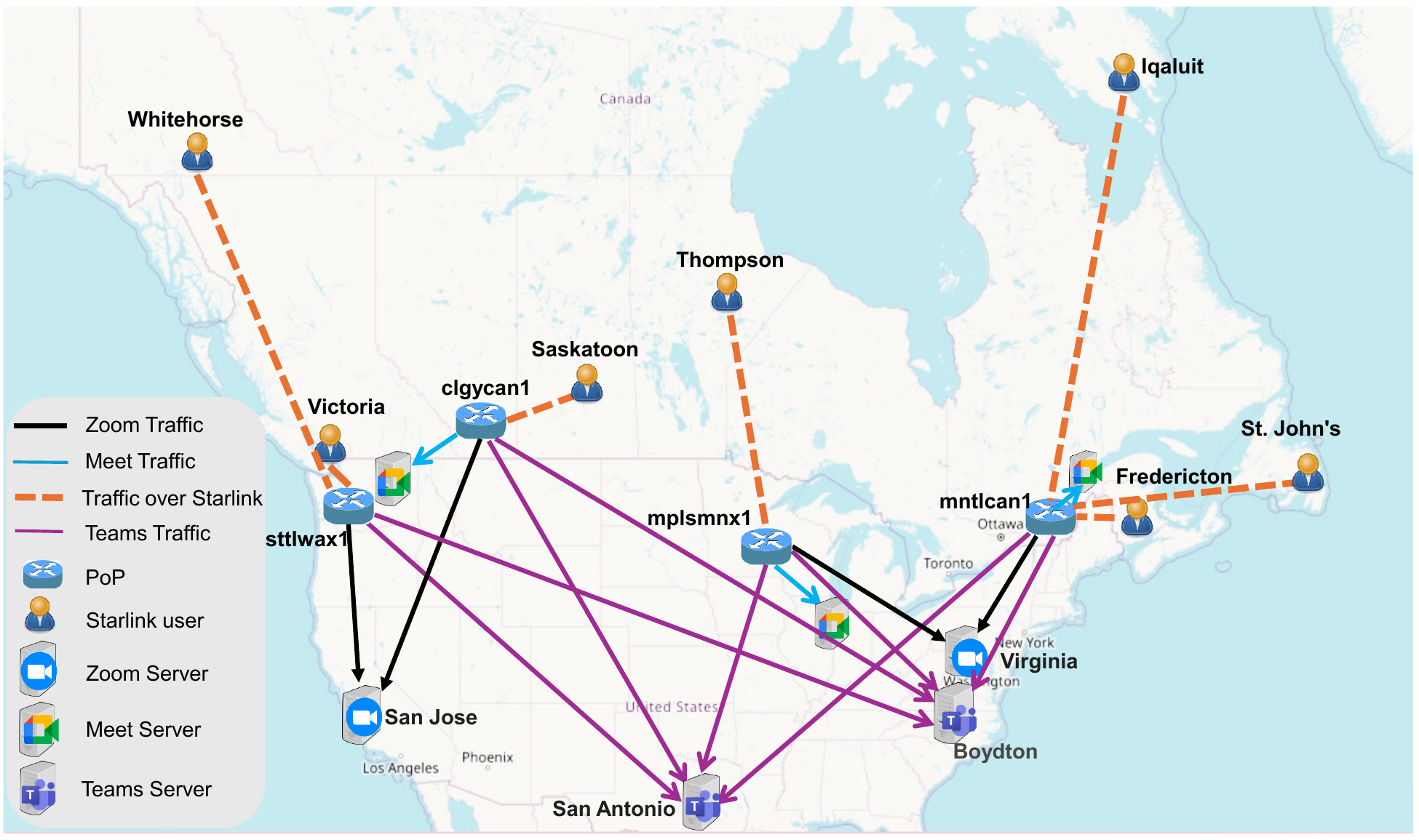}
    \caption{Overview of the seven users across Canada, their associated PoPs, and service points for Google Meet, Zoom, and Microsoft Teams.}
    \vspace{-3mm} 
    \label{fig:figLeo}
\end{figure}

{\bf 1) Service-Point Selection for Stationary Users:} For stationary Starlink users, service-point selection depends on both provider footprint and selection strategy. In our measurements, Google Meet service points are generally located near the associated Starlink PoP, consistent with broader observed edge coverage across our measurement locations. Zoom shows a more regional pattern, directing traffic from eastern Canadian PoPs mainly to Virginia and western PoPs to San Jose. Microsoft Teams is more centralized and session-anchored, often using a single service point regardless of user location, typically San Antonio or Boydton in our measurements. Consequently, latency differs across platforms and locations, with lower RTT usually observed when the selected service point is closer to the associated Starlink PoP. 

 {\bf 2) Service-Point Changes During PoP-Change Flights:} For in-flight users, platforms show different adaptation strategies during Starlink PoP changes. Microsoft Teams is largely session-anchored, keeping the same service point throughout a session despite PoP changes. Google Meet often shows service-point changes that are closely aligned with changes in the active Starlink access path. Zoom follows a coarser regional strategy, usually staying within a small set of regional hubs instead of adapting to every PoP change.

{\bf 3) Service-Point Stability and Peer Awareness:} When the Starlink PoP remains unchanged during a flight segment, Google Meet and Microsoft Teams generally keep stable service-point selections, with no changes in most flight segments. Zoom, however, sometimes switches service points even without an observed PoP handover, and these changes can coincide with higher RTT. More broadly, some observed service-point selections are not well aligned with latency or peer location. Google Meet and Zoom often choose service points near the Starlink PoP, but in cases such as Thompson--Victoria the selected paths are not well aligned with the peer location, resulting in hairpinning through Chicago for Google Meet and Virginia for Zoom. Microsoft Teams follows a session-anchored strategy and may keep users on a distant fixed service point, such as Boydton, even when a closer option such as San Jose exists. Overall, these results show that service-point selection over Starlink is provider-specific and can coincide with less favorable end-to-end paths, especially during in-flight sessions where the selected RTC service point is not well aligned with the current Starlink egress path or the peer location. To support reproducibility and facilitate follow-up measurements, we make our dataset publicly available.\footnote{\url{https://github.com/kmalekinasab/RTCMeasurement}}

 \begin{figure}[!t]
    \centering
    \includegraphics[width=\linewidth]{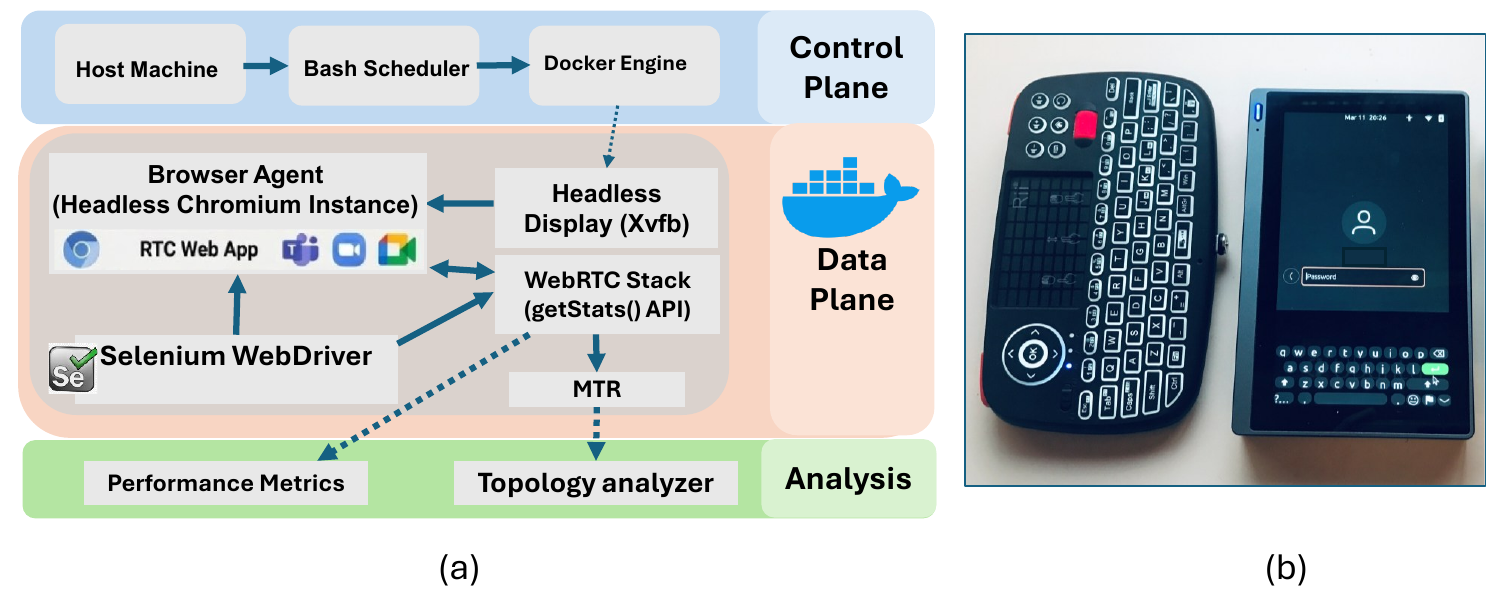}
     \vspace{-8mm}
    \caption{(a) The system setup used for both stationary and in-flight measurements. (b) A mini PC used during the flight measurements.}
    \label{fig:figtestbed}
\end{figure}

\par The rest of this paper is organized as follows. Section~\ref{Sec2} reviews the related work. Section~\ref{Measurement_objective} presents the measurement design. Sections~\ref{subsec:fixed-starlink-users} and~\ref{subsec:mobile-starlink-users} analyze service-point selection for stationary and in-flight Starlink users, respectively, and Section~\ref{sec:conclusion} concludes the paper. 
\section{Related Work}
\label{Sec2}

\par {\bf Measurement in terrestrial networks:} A passive packet-trace-based approach for measuring Zoom traffic in campus networks is proposed in \cite{michel2022enabling}. This methodology leverages observable, unencrypted header and traffic patterns to identify Zoom media streams and groups multiple RTP streams into discrete meetings. In addition, client-side measurements of Google Meet, Zoom, and Microsoft Teams across WiFi and 4G networks are presented by \cite{kumar2022comparison}. In this study, technical QoS metrics, such as data payloads and Inter-Packet Arrival Times, are analyzed alongside Quality of Experience (QoE) measures, including objective audio/video quality scores and subjective user feedback. Similarly, client-side measurements of the three major RTC providers under varying client-side conditions, such as video encoding, recovery times, differences in proprietary congestion control, and participants' viewing modes, are presented by \cite{macmillan2021measuring}. Additionally, the authors in \cite{chang2021can, chang2022tale} present a controlled measurement study of Zoom, Webex, and Google Meet, characterizing (i) their service infrastructure and endpoint architecture and (ii) user QoE, including streaming latency, video quality, traffic rate/bandwidth, and client resource usage. Using geo-distributed clients, the study compares performance across US regions (US-east vs. US-west) and Europe, and infers regional infrastructure effects primarily via RTT measurements. However, these prior measurements largely focus on client-side performance measurements under network- and browser-side configurations, or on the evaluation of major RTC providers across different client locations.

\par {\bf Measurement in LEO satellite networks:} The paper \cite{mohan2024multifaceted} compares the performance of Zoom over Starlink and terrestrial networks. The authors focused on one-way delay, packet loss, video bitrate/throughput, and frame rate. In addition, \cite{zhao2023realtime} studies the performance of screen sharing over Zoom on Starlink by measuring latency, jitter, and packet loss. However, most of the existing WebRTC measurements over Starlink focused on Zoom, and they considered stationary nodes in LEO satellite networks. Recent studies on in-flight Internet connectivity provided by LEO satellite networks \cite{jang2025geo, BorgianniLEO}  focus on network- and transport-layer behavior rather than application-layer RTC measurements. Recent Starlink and in-flight connectivity studies have characterized important aspects of LEO access networks, including egress evolution, latency dynamics, transport-layer behavior, and application-level quality of experience. Our study complements this line of work by shifting the focus from the Starlink access network alone to the interaction between Starlink egress dynamics and externally observable RTC service-point selection. Specifically, we examine how three major RTC platforms select provider-side service points for Starlink users, and how these selections vary across stationary and in-flight settings. This cross-provider perspective reveals behavior that is not visible from access-network measurements alone, including differences in PoP alignment, session anchoring, regional hub selection, and service-point stability during in-flight Starlink connectivity.

\section{Measurement Design Overview}
\label{Measurement_objective}

\par {\bf Objectives:}
RTC platforms such as Google Meet, Zoom, and Microsoft Teams are commonly deployed and optimized in terrestrial access networks. Over Starlink, however, the observed end-to-end path also depends on the Starlink PoP through which user traffic enters the terrestrial Internet. This PoP may differ across stationary locations and may change during flight. We therefore study how RTC platforms select service points over Starlink by evaluating: (i) service-point selection across fixed Starlink egress PoPs, (ii) service-point changes during in-flight PoP transitions, and (iii) service-point stability under a fixed PoP association.

\begin{figure}[t]
    \centering
    \includegraphics[width=\columnwidth]{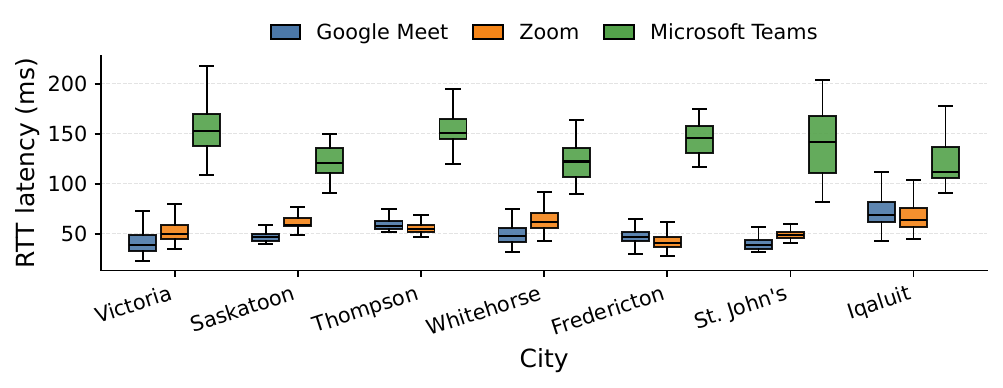}
    \caption{RTT distributions for stationary Starlink users across seven Canadian locations.}
    \label{fig:stationary-latency-boxplot}
    \vspace{-3mm}
\end{figure}

\par {\bf Measurement Methodology:}
To enable a controlled and repeatable comparison, we evaluate the web versions of Google Meet, Zoom, and Microsoft Teams. This choice is measurement-driven: WebRTC \texttt{getStats()} exposes the nominated Interactive Connectivity Establishment (ICE) candidate pair, RTT, candidate type, and remote candidate address needed to infer service-point selection and analyze paths through a consistent interface~\cite{webrtcstats}. We do not assume equivalence with native clients. Prior measurements show that browser and native versions can diverge, and that fine-grained native-client measurements are harder to obtain consistently~\cite{macmillan2021measuring}. Therefore, our results refer to the WebRTC-based versions of these providers. As shown in Figure~\ref{fig:figtestbed}, we automate sessions using Selenium and a custom JavaScript hook. For each active session, we identify the nominated ICE candidate pair exposed by WebRTC and extract the remote candidate address. We use this address as the starting point for service-point inference and define a service point as the externally visible provider-side RTC endpoint reached by the session. We then annotate each inferred service point with its ASN, operator, and approximate location using IP ownership and geolocation information, and compare it with the active Starlink egress PoP derived from path measurements.

We validate each service-point inference using multiple independent signals. We run \texttt{MTR} with both UDP- and TCP-based probes when available, examine path-level consistency between the Starlink egress PoP and the inferred service point, and compare observed addresses with publicly available provider endpoint information where applicable. We retain a service-point inference when the WebRTC candidate information, IP ownership, geolocation, and path measurements are consistent. We apply this validation procedure uniformly across Google Meet, Zoom, and Microsoft Teams, enabling a consistent comparison of externally visible provider-side service-point selection across stationary and in-flight Starlink sessions. Across our measurements, all platforms use UDP. Microsoft Teams uses relay-to-relay ICE pairs, whereas Google Meet and Zoom use non-relayed peer-reflexive-to-host pairs.

 \begin{figure*}[t]
    \centering
    \subfloat[Microsoft Teams map]{
        \includegraphics[width=0.3\linewidth]{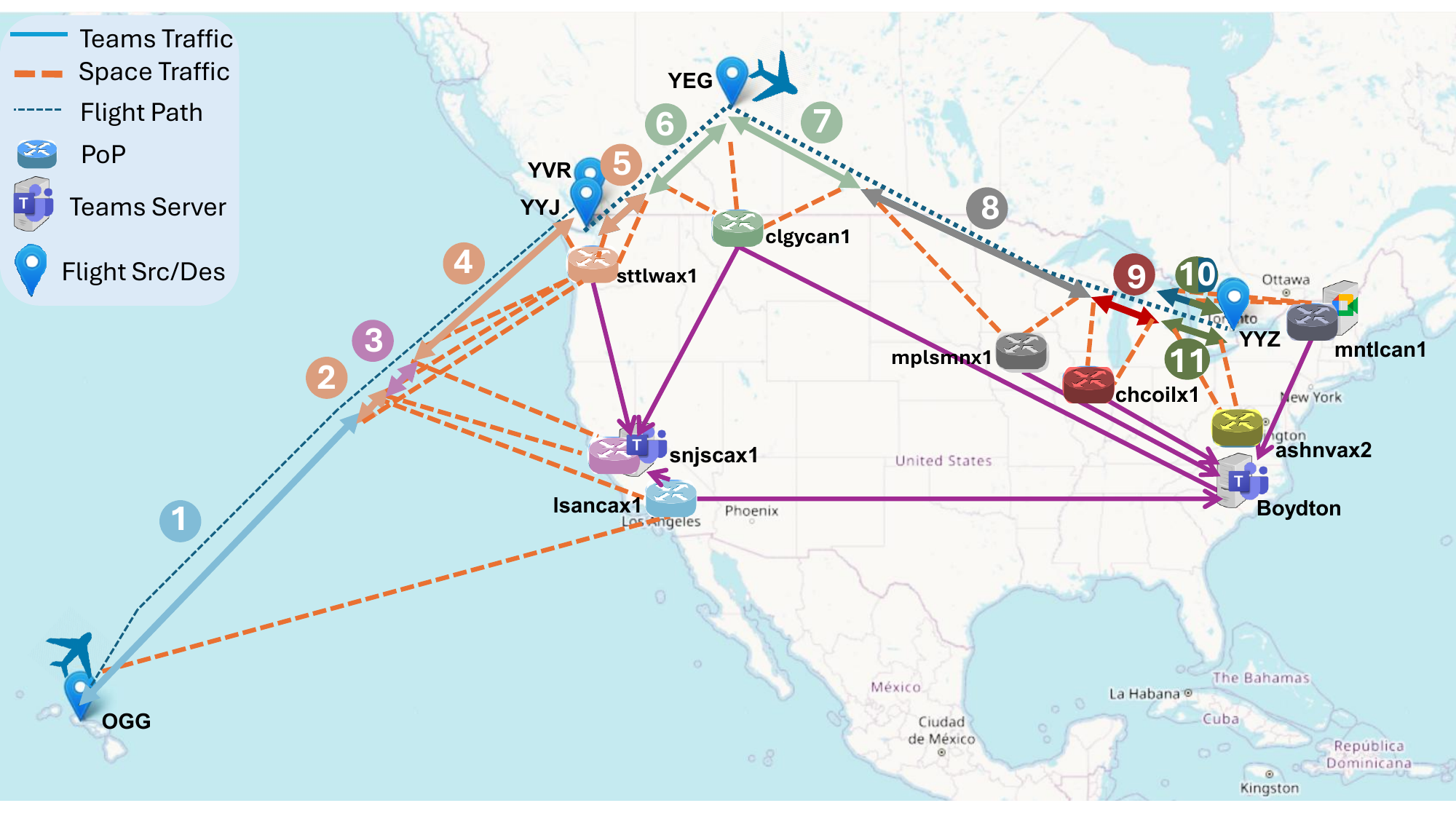}
        \label{fig:sub1}
    }
    \hfil 
    \subfloat[Google Meet map]{
        \includegraphics[width=0.3\linewidth]{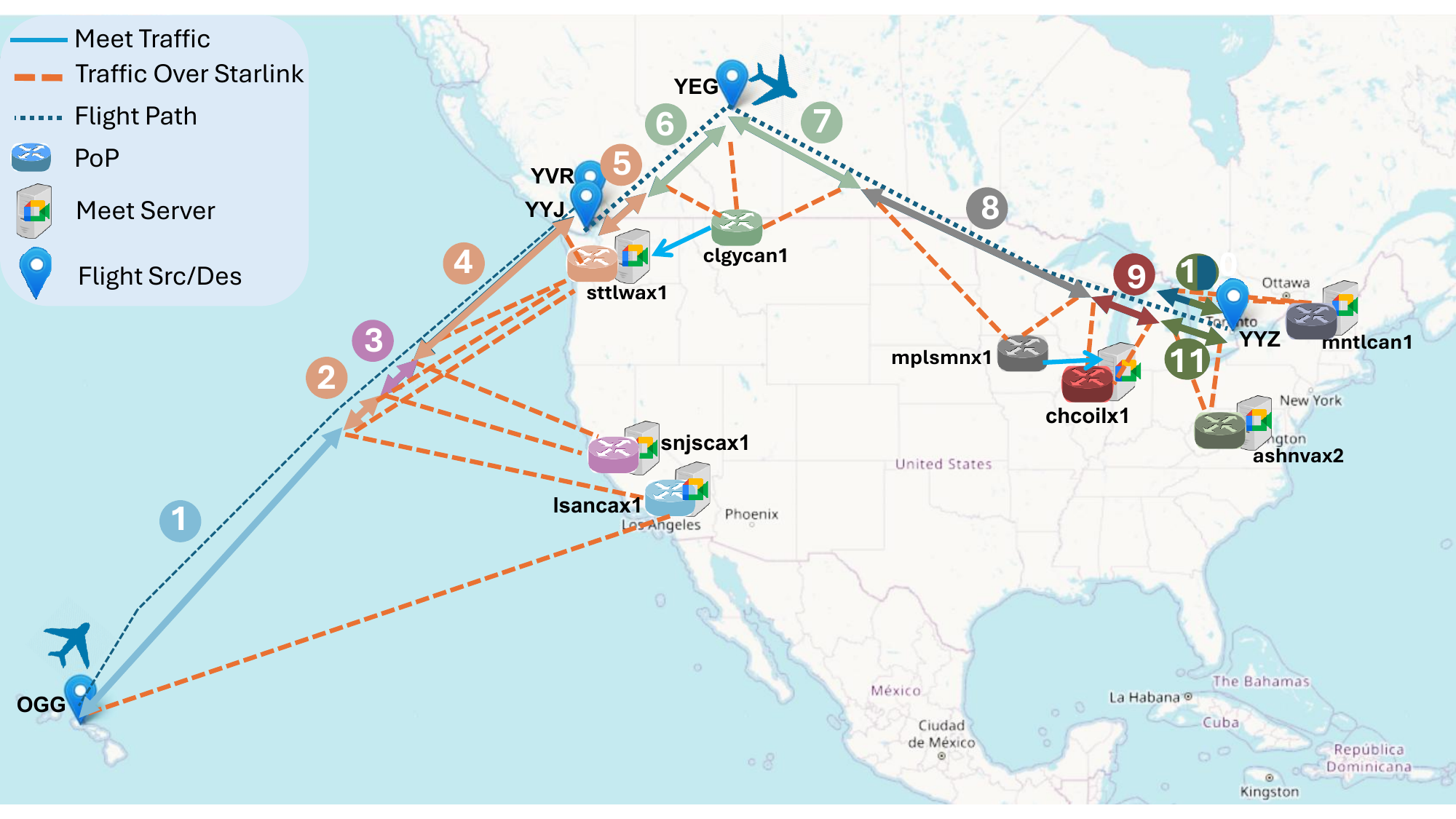}
        \label{fig:sub3}
    }
    \hfil
    \subfloat[Zoom map]{
        \includegraphics[width=0.3\linewidth]{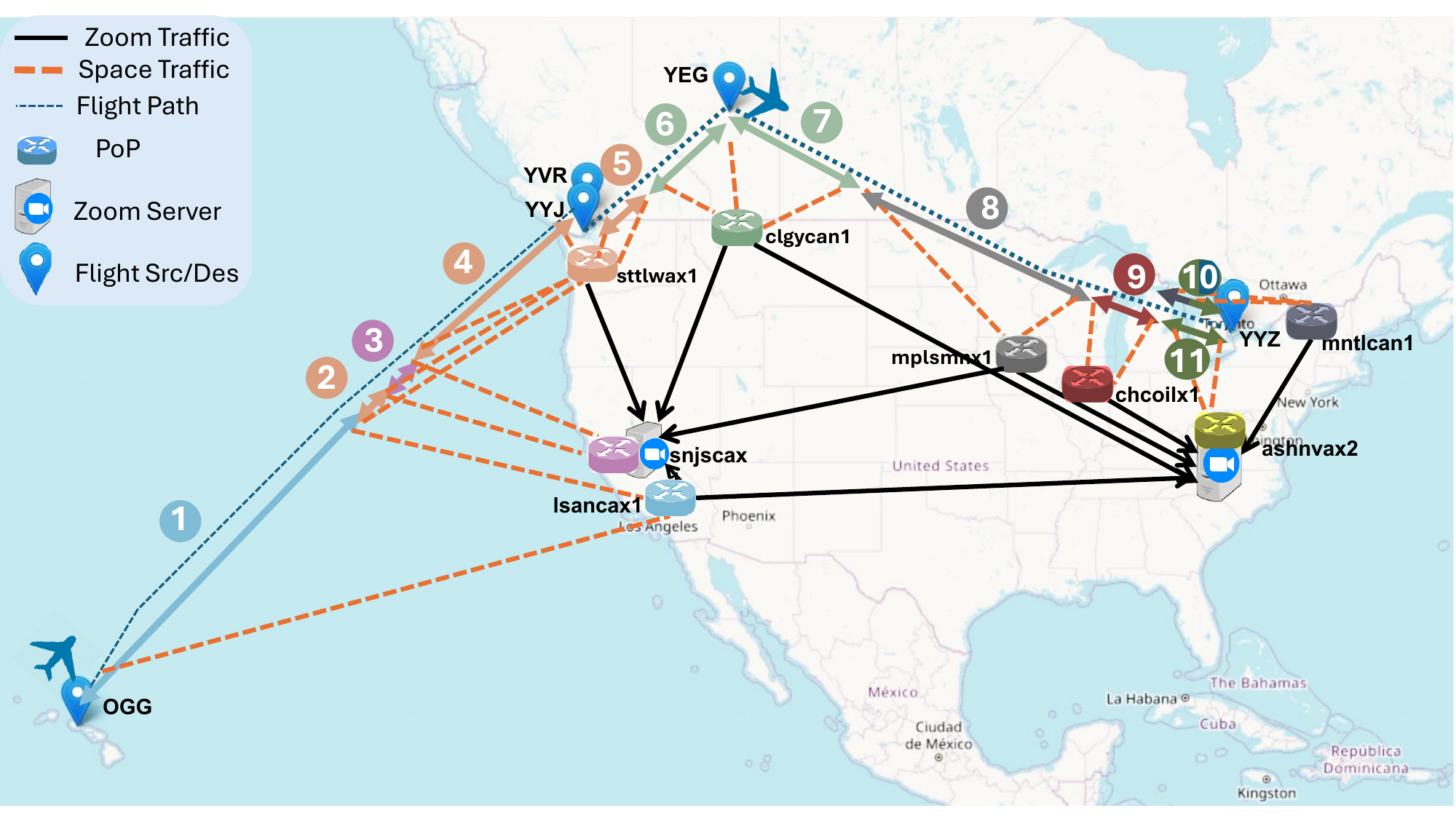}
        \label{fig:sub2}
    }

    \caption{Observed RTC traffic paths and service-point selection behavior over the Starlink LEO network. Subfigures (a)--(c) show the provider-specific routing behavior of Microsoft Teams, Google Meet, and Zoom.}
    \label{fig:main_figure1}
\end{figure*}

Our study covers both stationary and in-flight Starlink users. Stationary
measurements were conducted at seven Canadian locations: Victoria, Saskatoon,
Whitehorse, Iqaluit, Fredericton, St. John's, and Thompson, over multiple days
and time periods for more than 10 hours in total. In this setting, each user
terminal remained associated with a stable Starlink egress PoP during the
measurement period, allowing us to characterize service-point selection under
fixed egress conditions across geographically diverse PoPs. In-flight measurements were collected over six Starlink-enabled WestJet flights, totaling more than 24 hours across three round-trip routes: YYJ--YEG, YEG--YYZ, and YVR--OGG, with flight pairs WS502/WS509, WS456/WS467, and WS1850/WS1851, respectively. Rather than maintaining one continuous call for an entire flight, we repeatedly launched 10-minute sessions for each provider. This allows each session to
trigger a fresh service-point selection opportunity, helping us characterize the
provider's selection policy during flight: whether service points change with
Starlink egress PoPs, remain stable across sessions, or depend on factors beyond
the current PoP.
\section{Service-Point Selection for Stationary Starlink Users}
\label{subsec:fixed-starlink-users}

We analyze stationary Starlink users to characterize the externally visible RTC service points reached through Starlink egress paths under three settings: shared-PoP two-party sessions, different-PoP two-party sessions, and multi-party sessions. The Victoria node is included in all scenarios as a local control node to provide consistent session initiation and management.

\par {\noindent\bf 1) Two-Party Sessions With the Same PoP.} We first consider Victoria--Victoria and Victoria--Whitehorse sessions, where both endpoints are associated with the same Starlink PoP in Seattle (\texttt{sttlwax1}). Despite the same egress PoP, the three providers select different service points: Zoom routes traffic to San Jose, Microsoft Teams to San Antonio, and Google Meet to Seattle (see Figure~\ref{fig:figLeo}). This shows that provider-side service selection can shape the observed traffic path even when Starlink egress is identical.

\par {\noindent\bf 2) Two-Party Sessions With Different PoPs.} We next evaluate sessions between nodes associated with different Starlink PoPs: Victoria--Fredericton, Victoria--St. John's, Victoria--Thompson, Victoria--Saskatoon, and Victoria--Iqaluit. Under multi-PoP conditions, the platforms exhibit distinct service-point selection policies. Zoom follows a regional steering pattern: for Victoria--Fredericton, the eastern endpoint exits through Montreal (\texttt{mntlcan1}) and is served from Virginia, while Victoria exits through Seattle (\texttt{sttlwax1}) and is served from San Jose. Microsoft Teams shows low service-point diversity, consistently selecting San Antonio regardless of user location or PoP. Google Meet shows stronger observed alignment with Starlink egress geography: traffic from St. John's, Fredericton, and Iqaluit exits through \texttt{mntlcan1} and is served by Montreal-based infrastructure. For intermediate locations, Saskatoon exits through Calgary (\texttt{clgycan1}), where Teams, Google Meet, and Zoom select San Antonio, Seattle, and San Jose, respectively. Thompson exits through Minneapolis (\texttt{mplsmnx1}), with Google Meet selecting Chicago, Teams remaining anchored to San Antonio, and Zoom selecting Virginia. These results show that provider-side service deployment and selection behavior can lead to substantially different service-point assignments over the same Starlink access network.

\begin{figure}[t]
    \centering
    \includegraphics[width=\columnwidth]{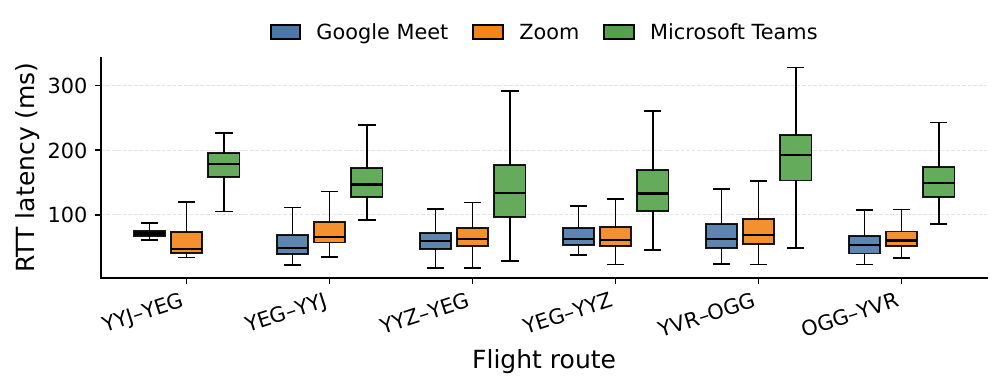}
    \caption{RTT distributions for Starlink-enabled in-flight RTC sessions.}
    \label{fig:flight-latency-boxplot}
\end{figure}

\par {\noindent\bf 3) Multi-Party Sessions.} Finally, we evaluate multi-party RTC sessions using three groups: Group~1 (Victoria, Whitehorse, Iqaluit), Group~2 (Fredericton, Thompson, Victoria), and Group~3 (Whitehorse, St.\ John's, Victoria). Microsoft Teams remains centralized: Groups~1 and~2 are routed to Boydton, while Group~3 is routed to San Antonio, indicating limited dependence on Starlink PoP proximity. Zoom preserves its regional behavior from the two-party case, with Iqaluit, Fredericton, and Thompson directed to Virginia, and Victoria and Whitehorse directed to San Jose. Google Meet shows the strongest observed PoP alignment: Victoria and Whitehorse exit through \texttt{sttlwax1} and connect to western regional servers, while Fredericton, Iqaluit, and St.\ John's exit through \texttt{mntlcan1} and connect to Montreal-based infrastructure. Thompson is an exception, exiting through \texttt{mplsmnx1} and connecting to Google's Chicago infrastructure. Overall, Google Meet is most closely aligned with the active Starlink egress PoP, while Teams is more centralized and Zoom follows a regional edge-cluster pattern.

\par {\noindent\bf RTT Impact of Service-Point Selection for Stationary Users.} Figure~\ref{fig:stationary-latency-boxplot} shows clear provider-specific
latency patterns for stationary Starlink users. Google Meet and Zoom generally
maintain low RTTs, with most samples below 75--100~ms and relatively stable
distributions. In contrast, Microsoft Teams consistently shifts toward higher
latency, with median RTTs around 120--160~ms and tails often extending beyond
180--200~ms. This gap suggests that Microsoft Teams' more centralized anchoring strategy is associated with higher latency, compared with the more localized service-point selection observed for Google Meet and Zoom. Beyond RTT, detailed inbound video statistics show that playback stability differs substantially across platforms. We report these city-level results in Appendix~\ref{app:inbound-video}, where we further analyze jitter variability, packet loss, freezes, bitrate adaptation, and codec behavior.
\begin{takeawaybox}
\noindent\textbf{Takeaway \#1:} {\bf For stationary Starlink users, service-point selection is highly provider-specific. Microsoft Teams anchors multi-party sessions to the service point selected for the first participant, so all participants use the same service point regardless of location. Zoom follows regional hub-based steering, while Google Meet is most closely aligned with the Starlink egress PoP and often selects nearby service points more consistently.}
\end{takeawaybox}
\vspace{-8pt} 
\section{In-flight Starlink Users}
\label{subsec:mobile-starlink-users}
This section examines RTC service-point selection in in-flight Starlink scenarios. Our goal is to understand how RTC platforms react when aircraft movement changes the Starlink egress PoP through which user traffic enters the terrestrial Internet. We evaluate whether these egress changes trigger service-point reselection, whether platforms preserve session anchoring despite PoP changes, and whether service points remain stable under the same PoP association.
\newcommand{\popid}[1]{\texttt{#1}\allowbreak}
\newcommand{\poparr}{\ensuremath{\allowbreak\rightarrow\allowbreak}}

\par { \bf Starlink PoP Trajectories During Flights.} As shown in
Figure~\ref{fig:main_figure1}, we first examine the sequence of
Starlink PoPs observed during each flight. On the
YYJ$\allowbreak\rightarrow\allowbreak$YEG flight, Starlink traffic is
routed through two PoPs, \popid{sttlwax1} and \popid{clgycan1}
(PoP in steps~\popnum{5} and~\popnum{6}), and the return flight
(YEG$\allowbreak\rightarrow\allowbreak$YYJ) uses the same PoPs in
reverse order (\popnum{6}\poparr\popnum{5}). On the
YEG$\allowbreak\rightarrow\allowbreak$YYZ flight, the PoP trajectory is
\popid{clgycan1}, \popid{mplsmnx1}, Chicago (\popid{chcoilx1}),
\popid{mntlcan1}, and Ashburn (\popid{ashnvax2})   (steps~\popnum{7}\poparr\popnum{8}\poparr\popnum{9}\poparr
\popnum{10}\poparr\popnum{11}), while the return flight follows a
slightly different path, resulting in a shortened reverse PoP sequence
through \popid{ashnvax2}, \popid{chcoilx1}, \popid{mplsmnx1}, and
\popid{clgycan1} (\popnum{11}\poparr\popnum{9}\poparr
\popnum{8}\poparr\popnum{7}). On the YVR--OGG flight, the trajectory is
PoPs~\popnum{4}\poparr\popnum{3}\poparr\popnum{2}\poparr\popnum{1}:
the traffic path was initially served through \popid{sttlwax1} before
transitioning to the San Jose PoP (\popid{snjecax1}). It subsequently
returned to Seattle and later stabilized at the Los Angeles PoP
(\popid{lsancax1}) for the remainder of the session. On the return leg,
traffic was routed through \popid{lsancax1} and \popid{sttlwax1}.

\begin{figure}[t]
    \centering
    \includegraphics[width=0.9\linewidth]{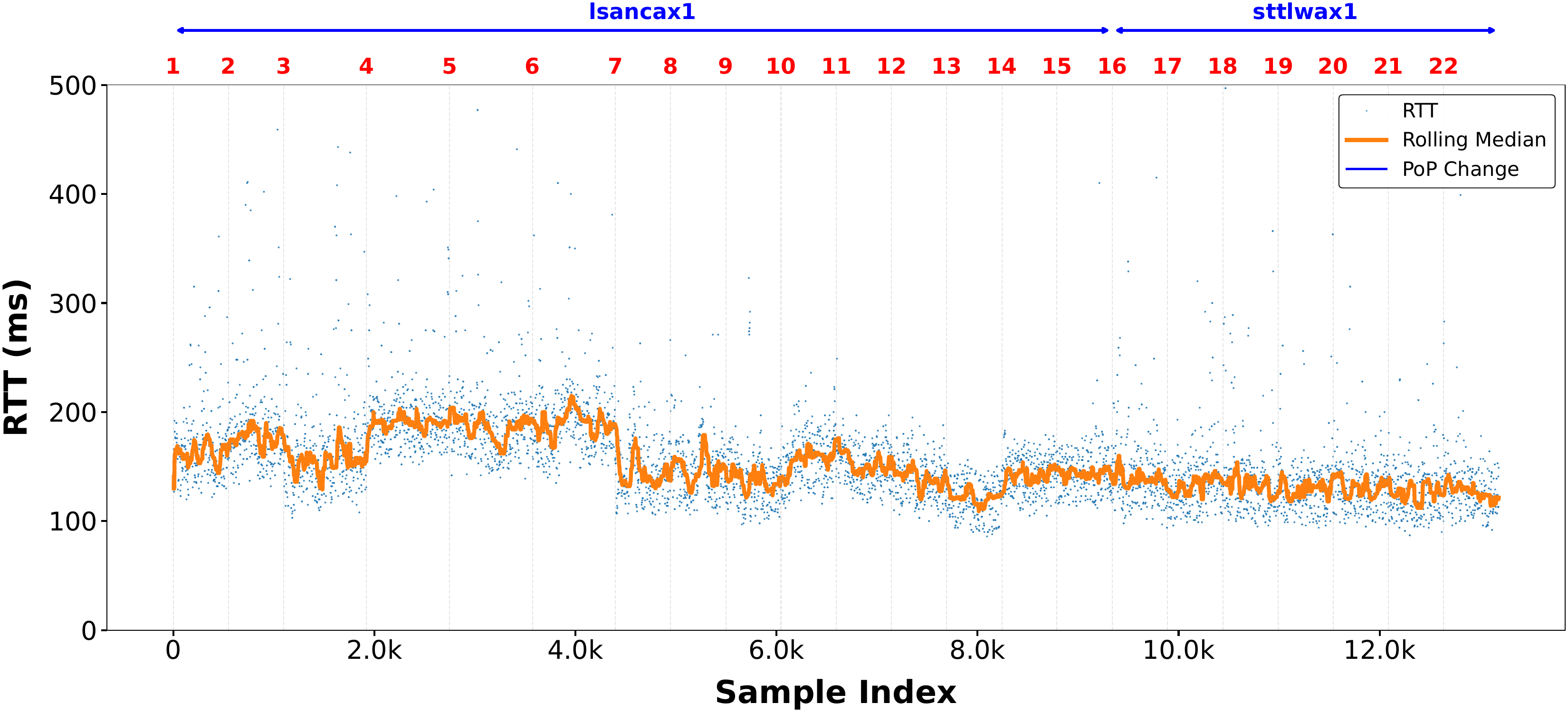}
    \caption{Microsoft Teams RTT dynamics during a Starlink-enabled flight on the
    OGG$\rightarrow$YVR route. } 
            \vspace{-3mm} 

    \label{fig:ms_teams_figs}
\end{figure}

\par {\bf A) Microsoft Teams -- Session-Anchored Service-Point Selection:}
Our in-flight measurements show that Microsoft Teams follows a
session-anchored service-point selection strategy: once a session is
established, the selected service point typically remains unchanged.
This behavior is visible across multiple flight routes. As shown in
Figure~\ref{fig:main_figure1}(a), traffic on the YYZ$\rightarrow$YEG route
remained anchored to the Boydton service point in both directions,
despite aircraft movement and underlying serving-PoP transitions. On the
YVR$\rightarrow$OGG route, the outbound session was anchored to San Jose,
whereas the return session was anchored to Boydton on the US east coast.
A similar pattern is observed on the YEG$\rightarrow$YYJ route, where the
outbound session consistently used the Virginia service point while the
return session was routed through San Jose; this route is examined
further in Appendix~\ref{appendix:additional-flights}
(Figure~\ref{fig:app_additional_flight_rtt}(a)).

Figure~\ref{fig:ms_teams_figs} further illustrates this behavior: even as
the associated serving PoP changes and RTT evolves over time, the
selected service point remains fixed. In this example, the meeting host
was located in Victoria while the aircraft moved toward the host, yet
the session traffic remained anchored to Boydton rather than shifting to
a service point closer to the users. This anchoring is also reflected in
the RTT distribution in Figure~\ref{fig:flight-latency-boxplot}, where
Microsoft Teams consistently exhibits higher RTTs than Google Meet and
Zoom across all in-flight routes, with several routes showing medians in
the 130--200~ms range and tails extending beyond 250--300~ms.

\begin{figure}[t]
    \centering
    \includegraphics[width=0.9\linewidth]{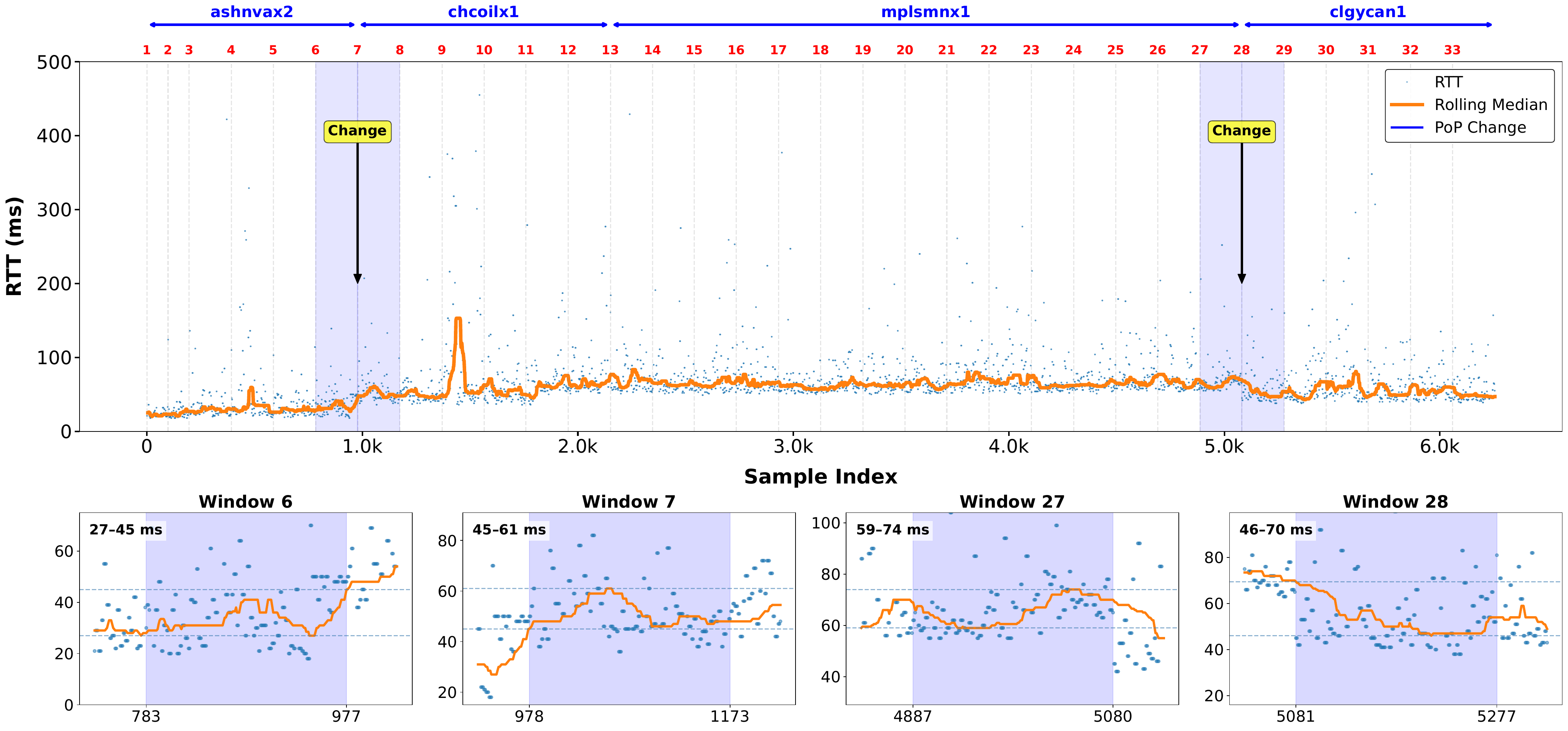}
    \caption{Google Meet RTT dynamics during a Starlink-enabled flight on the
    YYZ$\rightarrow$YEG route.}
            \vspace{-3mm} 

    \label{fig:google_meet_figs}
\end{figure}

\begin{figure*}[t]
    \centering
     \subfloat[YVR$\rightarrow$OGG]{
        \includegraphics[width=0.48\linewidth]{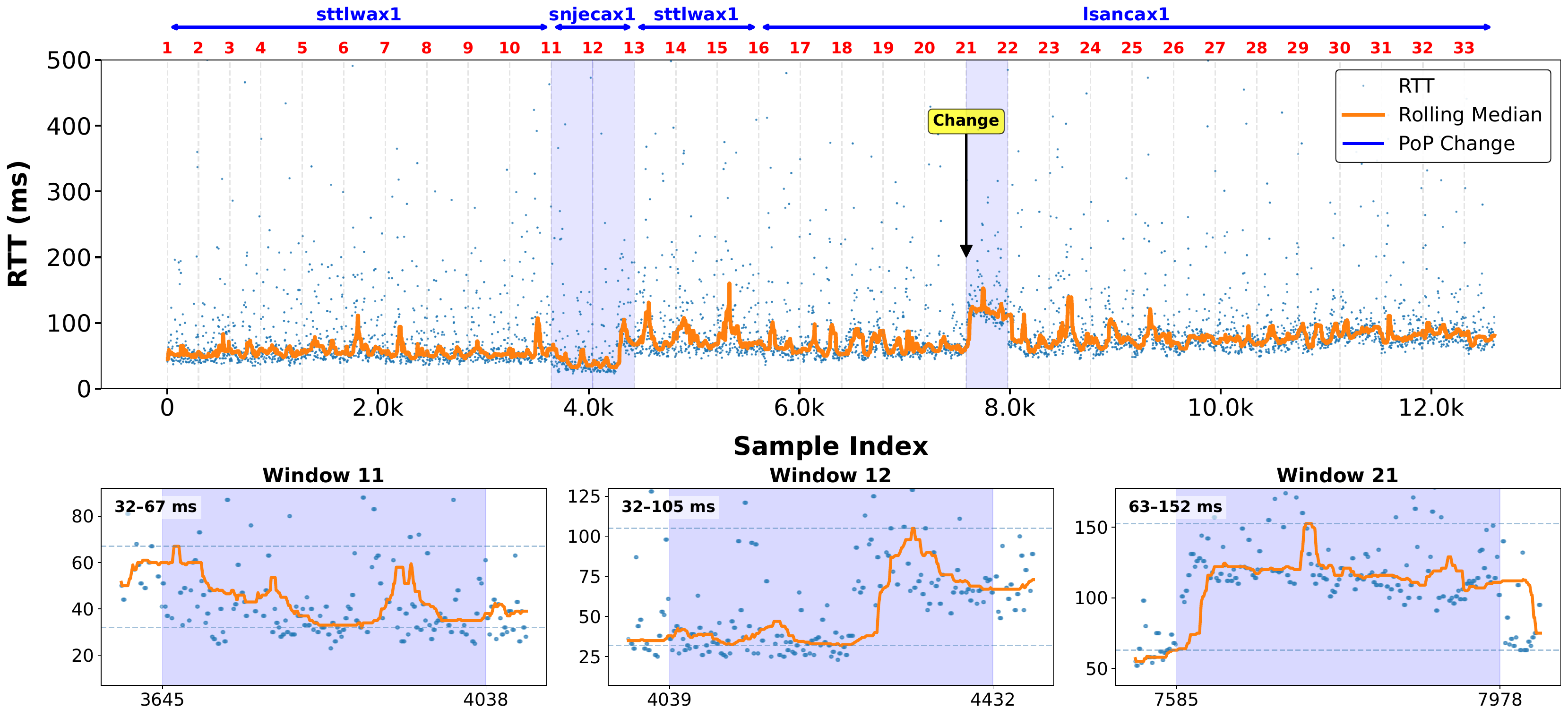}
        \label{fig:sub9}
    }
    \hfil
    \subfloat[YYZ$\rightarrow$YEG]{
        \includegraphics[width=0.48\linewidth]{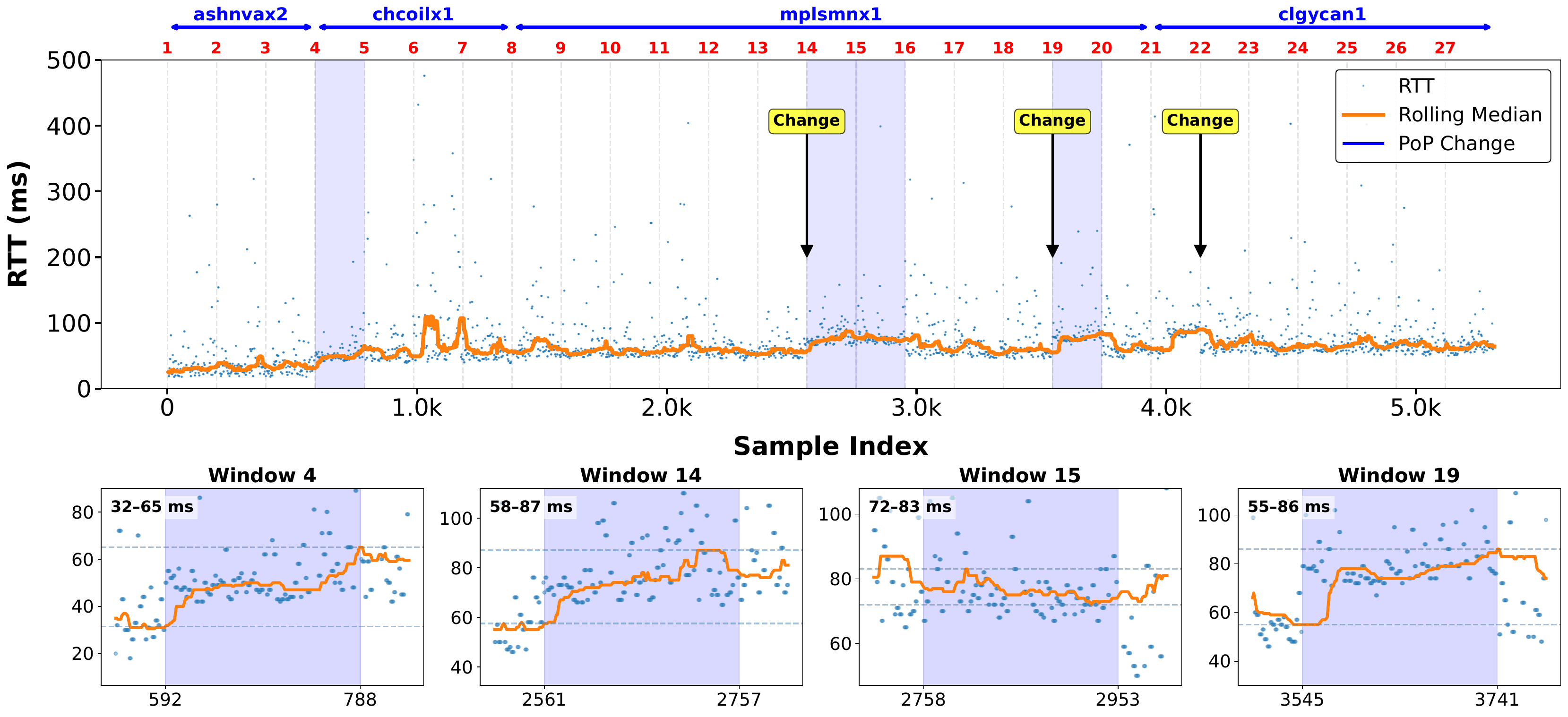}
        \label{fig:sub7}
    }
    \caption{Representative examples of Zoom RTT dynamics during Starlink-enabled flights on
    two routes. Each panel shows the RTT time series together with its rolling median.}
        \vspace{-2mm} 

    \label{fig:main_figure2}
\end{figure*}

\par {\bf B) Google Meet -- Dynamic Service-Point Selection Under Serving-PoP Transitions:}
\label{MEET-PArt}
Figure~\ref{fig:main_figure1}(b) shows that, unlike Microsoft Teams,
Google Meet exhibits dynamic service-point selection during flight. Path analysis shows that the observed Google Meet service point often changes as the associated Starlink serving PoP changes, and in most cases the selected service point is geographically nearby. Once selected
for a given PoP, the service point typically remains stable while the
session stays associated with that PoP. This behavior is particularly
clear on the YVR$\rightarrow$OGG route, where service-point changes are closely aligned with the observed serving-PoP transitions and the RTT evolution
is broadly consistent with aircraft movement relative to the associated
PoP; detailed measurements for this route are provided in
Appendix~\ref{appendix:additional-flights}
(Figure~\ref{fig:app_additional_flight_rtt}(b)).

However, this behavior is not uniform across all flights. On the
YYZ$\rightarrow$YEG route and its reverse direction, Google Meet does not
always select the geographically closest service point for intermediate
PoPs. For example, after the transition from \texttt{chcoilx1} to
\texttt{mplsmnx1}, the selected service point remains in Chicago, as
shown in Figure~\ref{fig:google_meet_figs}. A similar pattern is observed
for \texttt{clgycan1}, where Seattle is selected despite the availability
of closer service points
(Appendix~\ref{Appendix-service-point_selection}). These route-dependent exceptions are associated with higher RTT. One possible contributor is indirect AS-level connectivity between the Starlink PoP AS and Google's AS, which can introduce additional path stretch before traffic reaches the observed Google Meet service point
(Appendix~\ref{AS-Level-appendix}). Overall, the route-level RTT
distribution in Figure~\ref{fig:flight-latency-boxplot} shows that Google Meet
maintains a low-latency profile across flights, with compact RTT
distributions mostly below 100~ms and substantially lower RTTs than
Microsoft Teams, although some routes still show RTT increases when the
selected service point is not well aligned with the serving PoP.

\par {\bf C) Zoom -- Regional Hub Traffic Management with Instability:}
Figure~\ref{fig:main_figure1}(c) shows that Zoom follows a regional-hub
service-point selection strategy, where traffic is typically mapped to
one of two main U.S.\ clusters, San Jose or Virginia, rather than being
tightly coupled to the currently associated Starlink serving PoP.
However, this selection is less stable on some routes. On the
YVR$\rightarrow$OGG route, Zoom largely maintained a west-coast anchoring
strategy: outbound traffic stayed mostly in San Jose despite multiple
Starlink serving-PoP transitions, with only one brief switch to Virginia
before returning to San Jose. As shown in Figure~\ref{fig:main_figure2}(a),
after the transition to \texttt{snjecax1}, RTT dropped to a stable range
of 32--100~ms in sample windows 11--12, whereas a server re-routing event
in sample window 21 increased RTT to 152~ms, nearly doubling the
baseline.

On the YEG$\leftrightarrow$YYZ route, Zoom showed the most frequent service-point changes among the observed flights, switching between U.S.\ east- and west-coast service points at short intervals (Figure~\ref{fig:main_figure2}(b), Windows~14, 15, and~19). These changes occurred while the Starlink egress PoP remained fixed at \texttt{mplsmnx1}, suggesting that they were driven mainly by Zoom's steering rather than by Starlink PoP transitions. A PoP switch from \texttt{ashnvax2} to \texttt{chcoilx1} at Window~4 coincided with a noticeable RTT increase, showing that Starlink PoP handovers can also contribute to the larger RTT steps observed during the flight. Additional route examples for the YEG$\rightarrow$YYZ and YEG$\rightarrow$YYJ flights are provided in Appendix~\ref{appendix:additional-flights} (Figure~\ref{fig:app_additional_flight_rtt}(c)--(d)).


 Figure~\ref{fig:flight-latency-boxplot} shows that Zoom generally
maintains much lower RTTs than Microsoft Teams, but its RTT distributions are less compact than those of Google Meet on several routes. This wider spread is
consistent with Zoom's regional-hub behavior: when the selected hub
matches the serving PoP well, RTT remains low, but unnecessary
cross-region service-point switches introduce RTT inflation and longer
tails, especially on routes such as YYZ$\rightarrow$YEG and
YVR$\rightarrow$OGG.

\begin{takeawaybox}
\noindent{\bf Takeaway \#2:} {\bf During flights, the three providers follow different service-point selection strategies. Google Meet shows the strongest observed alignment with the Starlink egress region. Microsoft Teams remains session-anchored and can keep traffic on a fixed, distant service point such as Boydton even when other observed service points, such as San Jose, are geographically closer. Zoom follows a regional-hub strategy, typically using Virginia or San Jose rather than closely tracking each PoP transition. Under the same PoP association, Google Meet and Microsoft Teams generally keep stable service-point selections, whereas Zoom may change its routing even without a PoP handover. These Zoom service-point changes coincide with RTT variations, indicating that service-point stability and RTT stability can diverge during in-flight Starlink access.}
\vspace{-2mm}
 \end{takeawaybox}

\section{Conclusion}
\label{sec:conclusion}

This paper presented a measurement study of browser-based RTC over Starlink using a Canada-wide testbed and more than 24 hours of in-flight measurements. Our results show that Google Meet, Zoom, and Microsoft Teams expose different service-point selection behaviors over the same LEO access network. For stationary users, service-point choices are often stable but strongly provider-specific. During flights, Starlink egress PoP changes can shift the access path, while RTC providers respond differently: Google Meet often follows the serving region, Microsoft Teams is largely session-anchored, and Zoom follows regional hubs with occasional service-point changes that coincide with RTT inflation. These findings suggest that provider-side service-point selection and Starlink egress dynamics can interact in ways that affect latency, especially during in-flight connectivity. More broadly, our results motivate future RTC measurements that include terrestrial baselines, native clients, and controlled provider comparisons to separate LEO-specific effects from platform-specific infrastructure choices.

\bibliographystyle{ACM-Reference-Format} 
\bibliography{References.bib}

@misc{microsoftteams2023ecosystem,
author       = {Raghavan, Srini},
title        = {Year in Review: How {Microsoft Copilot}, {Microsoft Teams}, and Our Partners Built a Stronger Ecosystem},
howpublished = {\url{https://techcommunity.microsoft.com/blog/microsoftteamsblog/year-in-review-how-microsoft-copilot-microsoft-teams-and-our-partners-built-a-st/4013328}},
year         = {2023},
month        = dec,
note         = {Accessed: 2026-03-11}
}

@misc{zoom300m2021,
author       = {{Zoom}},
title        = {A Year Later: Reflecting and Looking Ahead},
howpublished = {\url{https://www.zoom.com/en/blog/reflecting-looking-ahead/}},
year         = {2021},
month        = mar,
note         = {Accessed: 2026-03-11}
}

@misc{Soltero2020Meet,
author       = {Soltero, Javier},
title        = {{Google Meet} Premium Video Meetings---Free for Everyone},
howpublished = {\url{https://blog.google/products/meet/bringing-google-meet-to-more-people/}},
year         = {2020},
month        = apr,
note         = {Accessed: 2025-10-27}
}

@inproceedings{mohan2024multifaceted,
  author    = {Mohan, Nitinder and Ferguson, Andrew E. and Cech, Hendrik and Bose, Rohan and Renatin, Prakita Rayyan and Marina, Mahesh K. and Ott, J{\"o}rg},
  title     = {A Multifaceted Look at Starlink Performance},
  booktitle = {Proceedings of the ACM Web Conference 2024},
  pages     = {2723--2734},
  year      = {2024},
  publisher = {Association for Computing Machinery},
  doi       = {10.1145/3589334.3645328}
}

@inproceedings{pan2023measuring,
title     = {Measuring a Low-Earth-Orbit Satellite Network},
author    = {Pan, Jianping and Zhao, Jinwei and Cai, Lin},
booktitle = {IEEE International Symposium on Personal, Indoor and Mobile Radio Communications (PIMRC)},
year      = {2023}
}

@inproceedings{michel2022enabling,
title     = {Enabling Passive Measurement of {Zoom} Performance in Production Networks},
author    = {Michel, Oliver and Sengupta, Satadal and Kim, Hyojoon and Netravali, Ravi and Rexford, Jennifer},
booktitle = {Proceedings of the ACM Internet Measurement Conference (IMC)},
pages     = {244--260},
year      = {2022}
}

@inproceedings{kumar2022comparison,
title     = {Comparison of Popular Video Conferencing Apps Using Client-Side Measurements on Different Backhaul Networks},
author    = {Kumar, Rohan and Nagpal, Dhruv and Naik, Vinayak and Chakraborty, Dipanjan},
booktitle = {Proceedings of the International Symposium on Theory, Algorithmic Foundations, and Protocol Design for Mobile Networks and Mobile Computing},
pages     = {241--246},
year      = {2022}
}

@inproceedings{macmillan2021measuring,
title     = {Measuring the Performance and Network Utilization of Popular Video Conferencing Applications},
author    = {MacMillan, Kyle and Mangla, Tarun and Saxon, James and Feamster, Nick},
booktitle = {Proceedings of the ACM Internet Measurement Conference (IMC)},
pages     = {229--244},
year      = {2021}
}

@inproceedings{chang2021can,
title     = {Can You See Me Now? A Measurement Study of {Zoom}, {Webex}, and {Meet}},
author    = {Chang, Hyunseok and Varvello, Matteo and Hao, Fang and Mukherjee, Sarit},
booktitle = {Proceedings of the ACM Internet Measurement Conference (IMC)},
pages     = {216--228},
year      = {2021}
}

@article{chang2022tale,
title   = {A Tale of Three Videoconferencing Applications: {Zoom}, {Webex}, and {Meet}},
author  = {Chang, Hyunseok and Varvello, Matteo and Hao, Fang and Mukherjee, Sarit},
journal = {IEEE/ACM Transactions on Networking},
volume  = {30},
number  = {5},
pages   = {2343--2358},
year    = {2022}
}

@inproceedings{jang2025geo,
title     = {From {GEO} to {LEO}: First Look Into {Starlink} In-Flight Connectivity},
author    = {Jang, HyunSeok Daniel and Varvello, Matteo and Raman, Aravindh and Zaki, Yasir},
booktitle = {Proceedings of the ACM Internet Measurement Conference (IMC)},
pages     = {952--962},
year      = {2025}
}

@inproceedings{BorgianniLEO,
title     = {Evaluating {LEO} Satellite Internet in Aeronautical Mobility: {A} Study of {Starlink} from {Hawaii} to {Japan}},
author    = {Borgianni, Luca and Adami, Davide and Giordano, Stefano},
booktitle = {IEEE International Conference on Computing, Networking and Communications (ICNC)},
year      = {2026}
}

@article{fang2025streaming,
title     = {Streaming Media over {LEO} Satellite Networking: A Measurement-Based Analysis and Optimization},
author    = {Fang, Hao and Zhao, Haoyuan and Wang, Feng and Chou, Yi Ching and Chen, Long and Shi, Jianxin and Liu, Jiangchuan},
journal   = {ACM Transactions on Multimedia Computing, Communications, and Applications},
volume    = {21},
number    = {9},
pages     = {1--24},
year      = {2025},
publisher = {ACM}
}

@inproceedings{zhao2023realtime,
title     = {Realtime Multimedia Services over {Starlink}: A Reality Check},
author    = {Zhao, Haoyuan and Fang, Hao and Wang, Feng and Liu, Jiangchuan},
booktitle = {Proceedings of the Workshop on Network and Operating System Support for Digital Audio and Video (NOSSDAV)},
pages     = {43--49},
year      = {2023}
}

@misc{alvestrand2021rfc,
author       = {Alvestrand, Harald T.},
title        = {Overview: Real-Time Protocols for Browser-Based Applications},
howpublished = {RFC 8825},
series       = {Request for Comments},
number       = {8825},
publisher    = {RFC Editor},
doi          = {10.17487/RFC8825},
url          = {https://www.rfc-editor.org/info/rfc8825},
year         = {2021}
}

@misc{BA_RTC,
author       = {{British Airways}},
title        = {{Wi-Fi Powered by Starlink}},
howpublished = {\url{https://www.britishairways.com/content/information/entertainment/wifi-on-board/starlink}},
note         = {Accessed: 2026-05-11}
}

@misc{Qatar_Starlink,
author       = {{Qatar Airways}},
title        = {{Starlink}: The Fastest On-Board Wi-Fi in the Sky},
howpublished = {\url{https://www.qatarairways.com/en/onboard/connectivity.html}},
note         = {Accessed: 2026-05-11}
}

@misc{webrtcstats,
author       = {{W3C Web Real-Time Communications Working Group}},
title        = {{Identifiers for WebRTC's Statistics API}},
howpublished = {\url{https://www.w3.org/TR/webrtc-stats/}},
note         = {W3C Candidate Recommendation Draft, 25 September 2025. Accessed: 2026-05-11}
}

\appendix

\begin{table*}[!htbp]
\centering
\renewcommand{\arraystretch}{1.15}
\setlength{\tabcolsep}{2.8pt}
\footnotesize

\begin{minipage}[!htbp]{0.49\textwidth}
    \centering
    \caption{Inbound video metrics for Google Meet}
    \label{tab:meet_inbound_compact}
    \begin{tabular}{lcccccccc}
        \toprule
        \textbf{City} & \textbf{Res.} & \textbf{FPS} & \textbf{BR} & \textbf{Jit $\mu\!\pm\!\sigma$} & \textbf{Jit rng} & \textbf{Loss} & \textbf{Frz} & \textbf{Dur} \\
        \midrule
        Fredericton & 720p & 29.31 & 1438.0 & 15.2 $\pm$ 5.3 & [5,88] & 0.12 & 24 & 8.76 \\
        Iqaluit     & 720p & 29.31 & 1434.5 & 12.4 $\pm$ 4.1 & [4,36] & 0.09 & 20 & 8.75 \\
        Saskatoon   & 720p & 29.31 & 1422.2 & 15.8 $\pm$ 4.6 & [5,57] & 0.11 & 24 & 10.25 \\
        St.\ John's & 720p & 29.29 & 1423.0 & 13.5 $\pm$ 4.0 & [5,42] & 0.05 & 32 & 10.88 \\
        Thompson    & 720p & 29.25 & 1430.2 & 13.1 $\pm$ 4.6 & [4,39] & 0.04 & 26 & 10.96 \\
        Whitehorse  & 720p & 29.08 & 1453.4 & 15.1 $\pm$ 5.5 & [6,65] & 0.07 & 34 & 12.97 \\
        \bottomrule
    \end{tabular}
\end{minipage}
\hfill
\begin{minipage}[!htbp]{0.49\textwidth}
    \centering
    \caption{Inbound video metrics for Microsoft Teams}
    \label{tab:teams_inbound_compact}
    \begin{tabular}{lcccccccc}
        \toprule
        \textbf{City} & \textbf{Res.} & \textbf{FPS} & \textbf{BR} & \textbf{Jit $\mu\!\pm\!\sigma$} & \textbf{Jit rng} & \textbf{Loss} & \textbf{Frz} & \textbf{Dur} \\
        \midrule
        Fredericton & 720p & 25.88 & 2062.9 & 9.6 $\pm$ 13.0 & [0,126] & 1.61 & 116 & 66.66 \\
        Iqaluit     & 720p & 27.53 & 2202.2 & 3.7 $\pm$ 3.7  & [1,71]  & 0.30 & 102 & 29.96 \\
        Saskatoon   & 720p & 28.85 & 2142.8 & 5.5 $\pm$ 6.0  & [0,84]  & 0.33 & 88  & 27.74 \\
        St.\ John's & 720p & 29.12 & 2200.8 & 6.2 $\pm$ 2.7  & [1,19]  & 0.15 & 89  & 25.44 \\
        Thompson    & 720p & 28.09 & 1992.4 & 8.4 $\pm$ 9.8  & [0,101] & 1.74 & 135 & 76.23 \\
        Whitehorse  & 720p & 27.62 & 2172.3 & 3.2 $\pm$ 2.1  & [0,14]  & 0.09 & 71  & 20.94 \\
        \bottomrule
    \end{tabular}
\end{minipage}

\vspace{0.2cm}
\begin{minipage}{\textwidth}
    \scriptsize
    \footnotesize
\textit{Note:} \textbf{Res.} denotes the dominant inbound video resolution over the session (pixels). \textbf{FPS} is the effective frame rate (frames/s). \textbf{BR} is the bit rate in kbps. \textbf{Jit} is reported as mean $\mu$, standard deviation $\sigma$, and range (ms). \textbf{Loss} is the packet loss in \%. \textbf{Frz} is the number of freeze events. \textbf{Dur} is the total freeze duration over a session of 600 s.
\end{minipage}
\end{table*}

\section{Inbound Video Performance Details}
\label{app:inbound-video}


Tables~\ref{tab:meet_inbound_compact} and~\ref{tab:teams_inbound_compact} provide detailed inbound video statistics for Google Meet and Microsoft Teams. The results show that jitter variability, rather than mean jitter alone, is an important differentiator. Microsoft Teams often exhibits wider jitter fluctuations across cities. For example, in Fredericton, Thompson, and Saskatoon, its jitter ranges are [0,126]~ms, [0,101]~ms, and [0,84]~ms, respectively, compared with [5,88]~ms, [4,39]~ms, and [5,57]~ms for Google Meet. These larger excursions coincide with higher packet-loss rates, more freeze events, and longer total freeze duration, suggesting that occasional severe jitter spikes can degrade playback stability.

This difference is notable because both platforms operate around a comparable 720p inbound resolution profile across the common cities. Despite this similarity, Google Meet maintains a lower inbound bitrate, higher frame rate, and fewer playback impairments. Sender-side logs further indicate different adaptation behaviors: Google Meet uses VP9 with a stable encoding profile and \texttt{\seqsplit{qualityLimitationReason}} remaining \texttt{none}, whereas Microsoft Teams uses H.264 and frequently reports \texttt{bandwidth} limitation. Overall, Google Meet provides more stable playback, while Microsoft Teams often delivers higher video quality with more packet loss, freeze events, and instability under changing network conditions.

Unlike Google Meet and Microsoft Teams, Zoom exposes limited browser-accessible telemetry. It provides ICE-related information and RTT to the selected peer candidate, but not detailed video-quality metrics such as jitter, frame rate, resolution, or bitrate.

\section{Additional In-Flight RTT Measurements}
\label{appendix:additional-flights}

This appendix presents additional route-specific measurements for Microsoft Teams, Google Meet, and Zoom that complement the main-text analysis.

\begin{figure*}[t]
    \centering
    \subfloat[Microsoft Teams: YEG$\rightarrow$YYJ]{
        \includegraphics[width=0.48\linewidth]{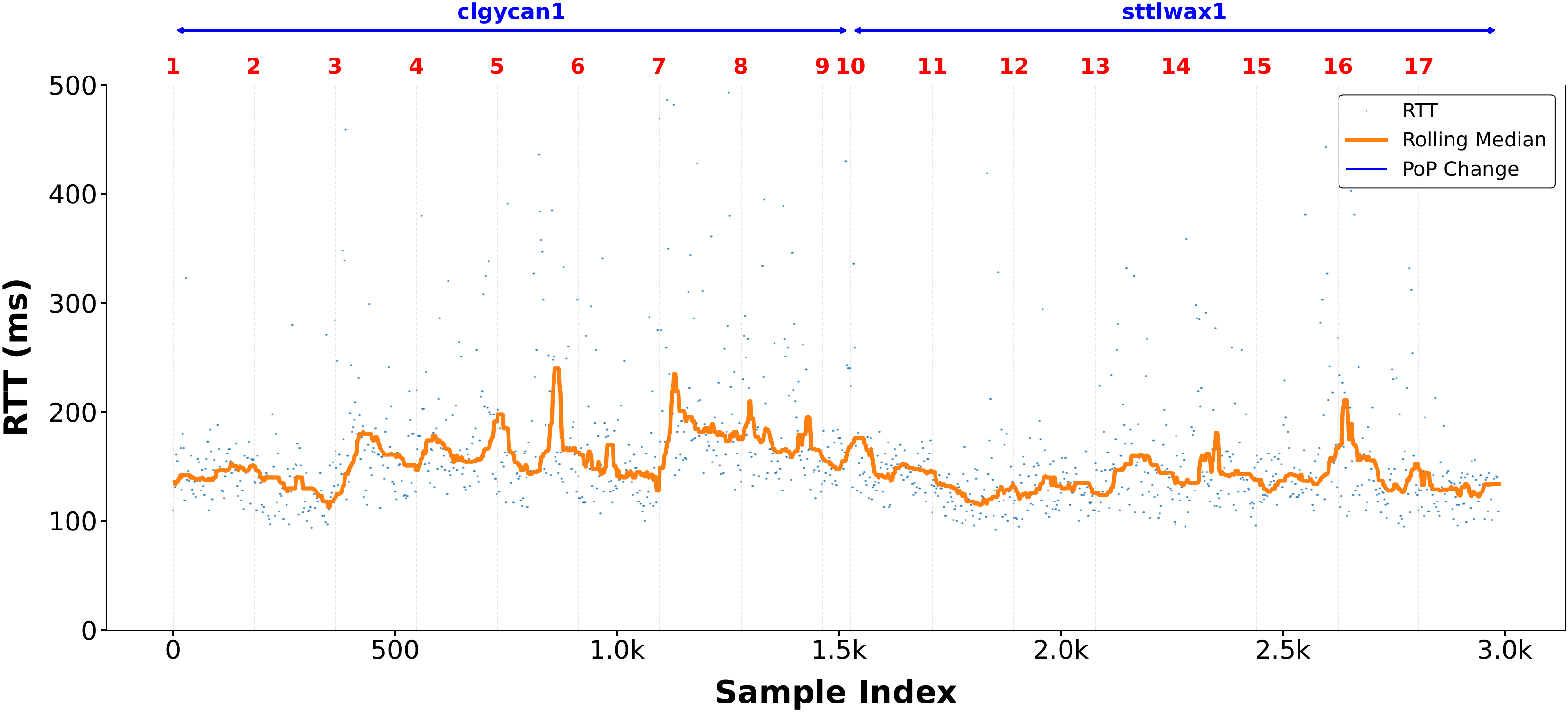}
        \label{fig:app_teams_edm_vic}
    }
    \hfil
    \subfloat[Google Meet: YVR$\rightarrow$OGG]{
        \includegraphics[width=0.48\linewidth]{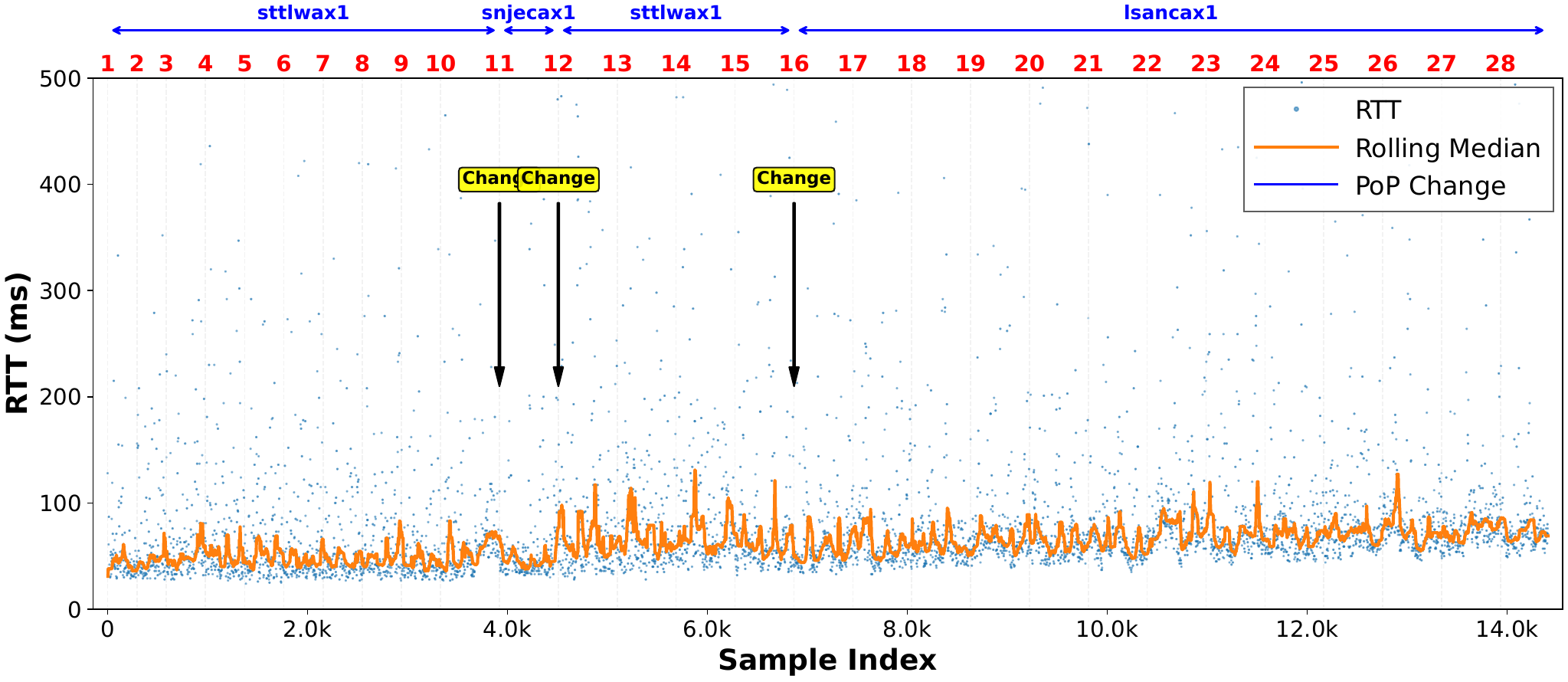}
        \label{fig:app_meet_van_hawaii}
    }
    \vspace{0.3cm}
    \subfloat[Zoom: YEG$\rightarrow$YYZ]{
        \includegraphics[width=0.48\linewidth]{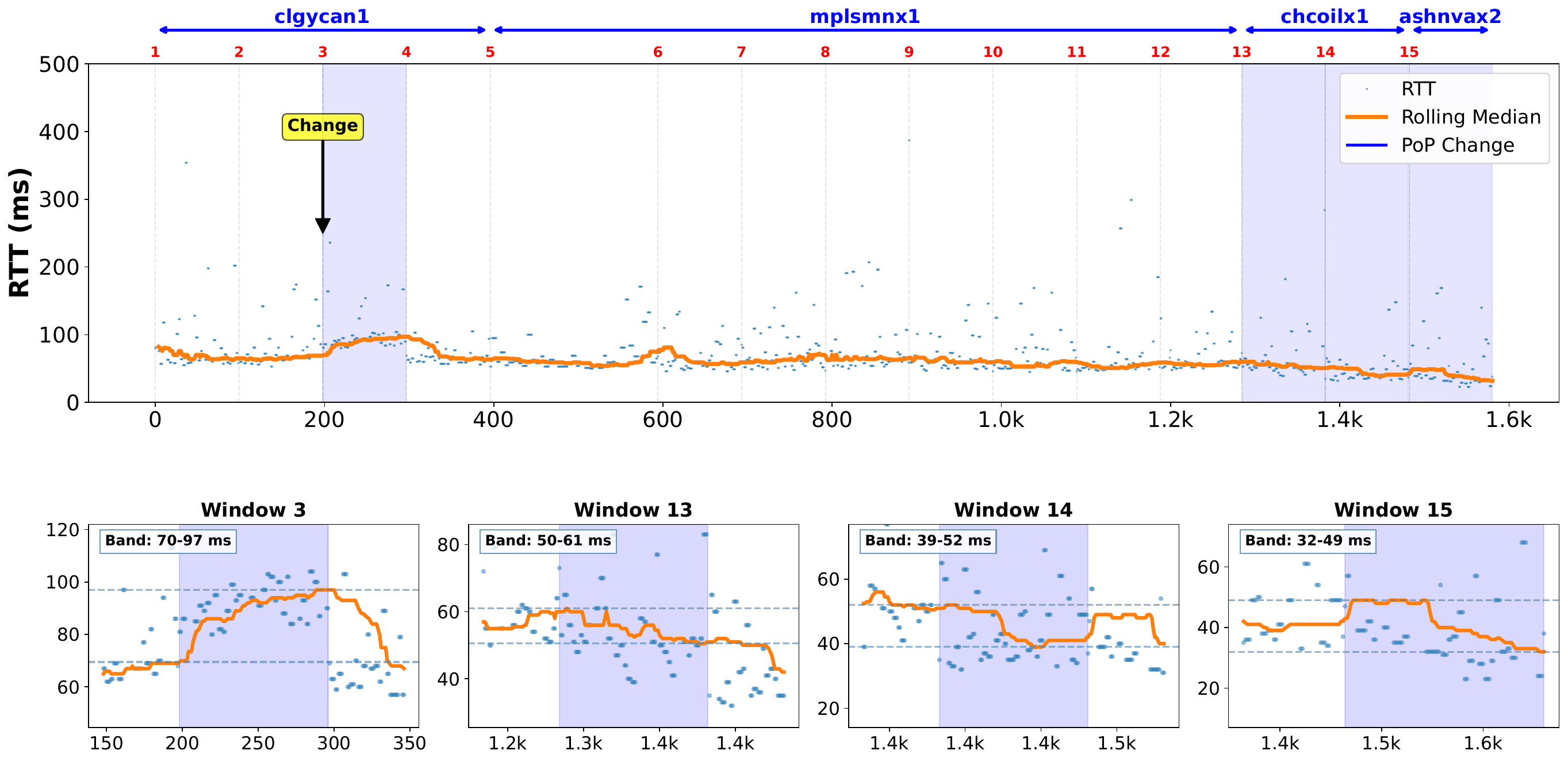}
        \label{fig:app_zoom_edm_tor}
    }
    \hfil
    \subfloat[Zoom: YEG$\rightarrow$YYJ]{
        \includegraphics[width=0.48\linewidth]{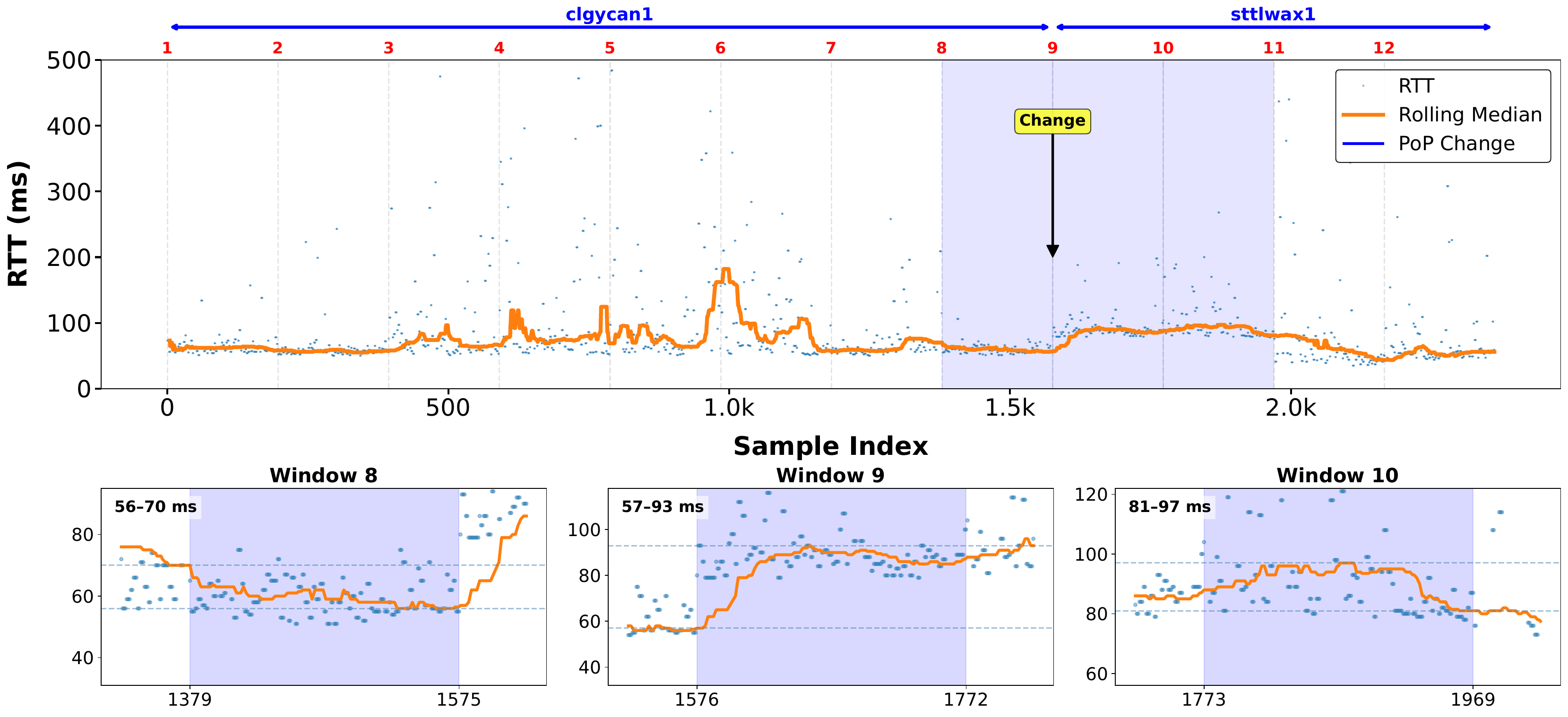}
        \label{fig:app_zoom_edm_vic}
    }
    \caption{Additional RTT dynamics during Starlink-enabled flights for Microsoft Teams, Google Meet, and Zoom. These examples complement the main-text figures by showing route-specific behavior across different flight directions and Starlink PoP transitions.}
    \label{fig:app_additional_flight_rtt}
\end{figure*}

\paragraph{\bf Microsoft Teams -- YEG$\rightarrow$YYJ}
A similar session-anchored pattern to that described in Section~\ref{subsec:mobile-starlink-users} is observed on the YEG$\rightarrow$YYJ route, as shown in Figure~\ref{fig:app_additional_flight_rtt}(a). The outbound session consistently used the Virginia service point, while the return session was routed through San Jose. This path selection remained stable and showed no clear correlation with RTT fluctuations, the direction of traffic toward the host, or changes in the available Starlink PoPs.

\paragraph{\bf Google Meet -- YVR$\rightarrow$OGG}
The Google Meet behavior discussed in Section~\ref{MEET-PArt} is particularly clear on the YVR$\rightarrow$OGG route and in the reverse direction, where service-point changes closely follow the observed PoP transitions, as shown in Figure~\ref{fig:app_additional_flight_rtt}(b). Google Meet shifts its selected service points in step with the Starlink PoP changes, and the RTT evolution is broadly consistent with the aircraft moving closer to or farther from the currently associated PoP. During each period in which the user remains associated with the same PoP, the selected service point remains stable, indicating that Google Meet does not show service-point changes while the Starlink egress remains unchanged.

\paragraph{\bf Zoom -- YEG$\rightarrow$YYZ}
On the YEG$\rightarrow$YYZ route, traffic was initially routed to San Jose. Before Starlink fully transitioned from \texttt{clgycan1} to \texttt{mplsmnx1}, Zoom traffic oscillated between San Jose and Virginia, as shown in Figure~\ref{fig:app_additional_flight_rtt}(c), sample window 3, and then returned to San Jose. Once the session stabilized on \texttt{mplsmnx1}, the observed traffic was persistently associated with Virginia through \texttt{chcoilx1} and \texttt{mntlcan1} in sample windows 13--15, reducing RTT to a stable range of 32--49~ms by the end of the flight.

\paragraph{\bf Zoom -- YEG$\rightarrow$YYJ}
On the YEG$\rightarrow$YYJ flights, Zoom traffic was mostly routed to San Jose, as shown in Figure~\ref{fig:app_additional_flight_rtt}(d). Only during a short interval after association with \texttt{clgycan1} did traffic briefly switch to Virginia before returning to San Jose. On the return flight, during the \texttt{clgycan1}-to-\texttt{sttlwax1} PoP transition, traffic again temporarily moved to Virginia and then returned to San Jose.

\section{Direct and Indirect AS-Level Connectivity from Starlink to Google}
\label{AS-Level-appendix}

Our analysis shows that direct AS-level connectivity from Starlink to Google is common on the measured terrestrial flight routes, but not universal. In particular, the YYZ$\rightarrow$YEG and YEG$\rightarrow$YYJ flights exhibit both direct and indirect connectivity depending on the active Starlink PoP. Table~\ref{tab:pop_google_patterns} summarizes these cases by listing, for each observed PoP, whether the path from Starlink AS14593 to Google AS15169 is direct or passes through one or more transit ASes. Similar behavior is also observed on the return flights, YEG$\rightarrow$YYZ and YYJ$\rightarrow$YEG.

This distinction is associated with a clear RTT effect. As shown in Table~\ref{tab:direct_indirect_rtt}, direct connectivity yields lower average RTT, whereas indirect connectivity increases RTT because traffic traverses intermediate transit ASes before reaching Google's network. In contrast, the long oceanic YVR$\rightarrow$OGG route consistently exhibits direct AS-level connectivity in our measurements, as shown in Table~\ref{tab:pop_google_patterns}, yet RTT remains substantially higher. This suggests that direct entry into Google's network helps reduce latency on terrestrial routes, whereas on the long oceanic route RTT can remain high because traffic may still traverse longer terrestrial or submarine paths before reaching the selected Google service point.

\begin{table}[!htbp]
\centering
\caption{Observed direct and indirect AS-level connectivity patterns from Starlink PoPs to Google's network. A pattern is classified as direct when traffic moves from AS14593 (Starlink) directly into AS15169 (Google), and indirect when one or more transit ASes appear in between.}
\label{tab:pop_google_patterns}
\scriptsize
\renewcommand{\arraystretch}{1.1}
\setlength{\tabcolsep}{3pt}
\begin{tabular}{|l|l|l|p{3.5cm}|}
\hline
\textbf{Flight} & \textbf{PoP} & \textbf{Pattern} & \textbf{Example} \\
\hline
YYZ$\rightarrow$YEG & \texttt{ashnvax2} & Direct & AS14593 $\rightarrow$ AS15169 \\
\hline
YYZ$\rightarrow$YEG & \texttt{chcoilx1} & Direct & AS14593 $\rightarrow$ AS15169 \\
\hline
YYZ$\rightarrow$YEG & \texttt{mplsmnx1} & Indirect & \shortstack[l]{AS14593 $\rightarrow$ AS3257 $\rightarrow$ AS1299 \\ $\rightarrow$ AS15169} \\
\hline
YYZ$\rightarrow$YEG & \texttt{clgycan1} & Indirect & \shortstack[l]{AS14593 $\rightarrow$ AS6461 $\rightarrow$ AS6327 \\ $\rightarrow$ AS15169} \\
\hline
YEG$\rightarrow$YYJ & \texttt{clgycan1} & Indirect & AS14593 $\rightarrow$ AS6327 $\rightarrow$ AS15169 \\
\hline
YEG$\rightarrow$YYJ & \texttt{sttlwax1} & Direct & AS14593 $\rightarrow$ AS15169 \\
\hline
YVR$\rightarrow$OGG & \texttt{sttlwax1} & Direct & AS14593 $\rightarrow$ AS15169 \\
\hline
YVR$\rightarrow$OGG & \texttt{snjecax1} & Direct & AS14593 $\rightarrow$ AS15169 \\
\hline
YVR$\rightarrow$OGG & \texttt{lsancax1} & Direct & AS14593 $\rightarrow$ AS15169 \\
\hline
\end{tabular}
\end{table}

\begin{table}[!t]
\centering
\caption{Average Google Meet RTT under direct and indirect AS-level connectivity patterns.}
\label{tab:direct_indirect_rtt}
\scriptsize
\renewcommand{\arraystretch}{1.1}
\setlength{\tabcolsep}{4pt}
\begin{tabular}{|l|c|c|c|}
\hline
\textbf{Flight} & \textbf{Direct RTT (ms)} & \textbf{Indirect RTT (ms)} & \textbf{Difference (ms)} \\
\hline
YYZ$\rightarrow$YEG & 54.9 & 71.4 & +16.5 \\
\hline
YEG$\rightarrow$YYJ & 59.1 & 70.3 & +11.1 \\
\hline
YVR$\rightarrow$OGG & 81.7 & N/A & N/A \\
\hline
\end{tabular}
\end{table}

\section{Proximity Mismatch in Service-Point Selection}
\label{Appendix-service-point_selection}

\begin{figure}[t]
    \centering
    \includegraphics[width=0.95\columnwidth]{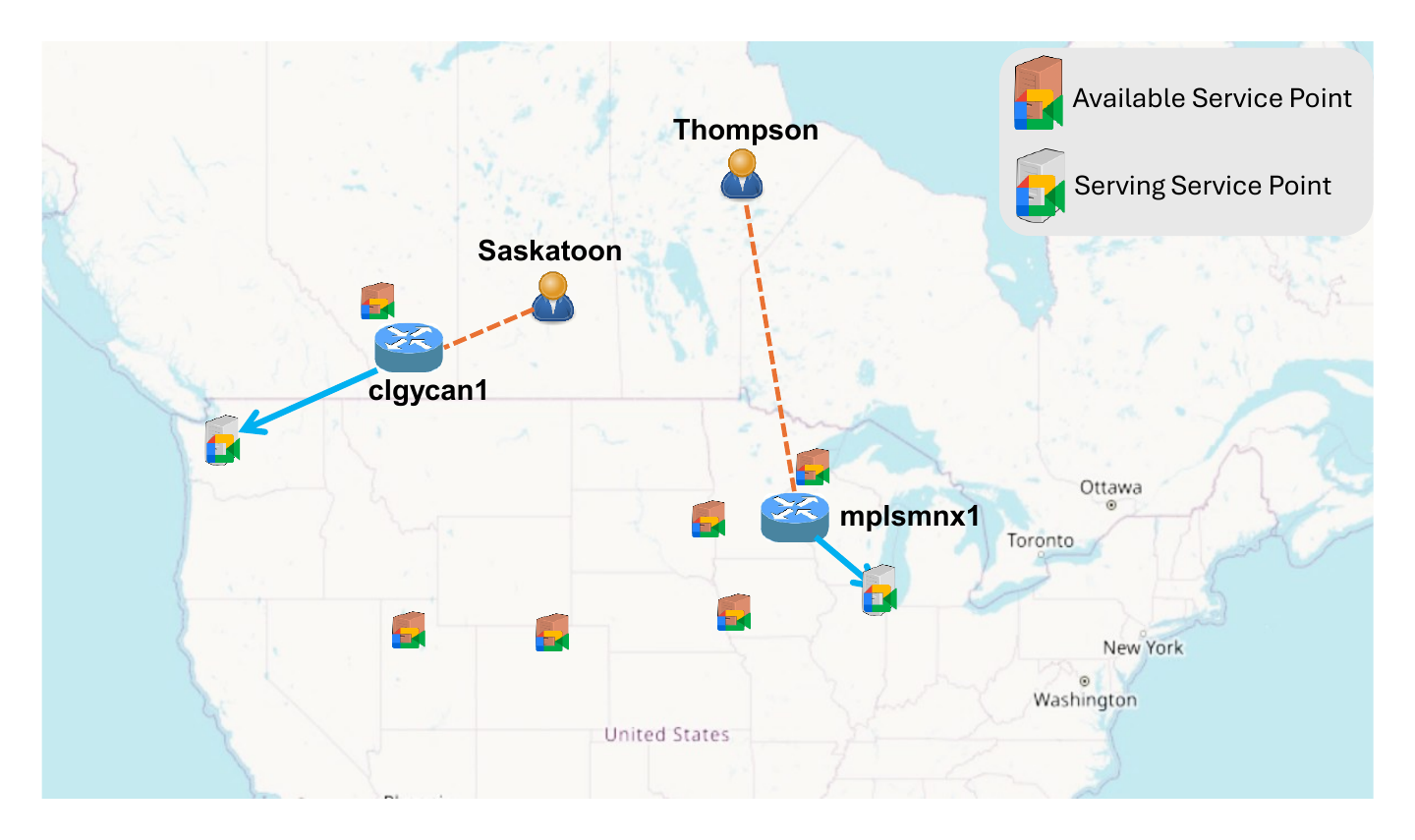}
    \caption{Observed Google Meet service-point selection for users in Saskatoon and Thompson.}
    \label{fig:meet_associate_pop}
\end{figure}

Several Starlink PoPs, including Calgary and Minneapolis, are observed to associate with Google service points in Seattle and Chicago, respectively, even though other Google service points appear to be geographically closer to those PoPs, as shown in Figure~\ref{fig:meet_associate_pop}. This suggests that Google Meet does not always select the nearest available service point from the Starlink egress location. Instead, the observed selection may reflect backend constraints or service-assignment factors that are not visible from our measurements. As a result, traffic may be directed to farther service points despite the presence of closer alternatives, indicating that service-point selection is not strictly proximity-aware.

\end{document}